\documentclass{rspublic}
\usepackage{graphicx}
\usepackage{amsbsy}

\begin{document}

\newcommand{\ea}{{\it et al. }}

\title{Electron-electron and electron-hole pairing in graphene structures}

\author[Yu.E. Lozovik, S.L. Ogarkov \& A.A. Sokolik]{Yu.\,E. Lozovik$^1$, S.L. Ogarkov$^2$ and A.A. Sokolik$^1$}

\affiliation{${}^1$ Institute of Spectroscopy, Russian Academy of Sciences, 142190 Troitsk, Moscow Region, Russia\\
${}^2$ National Nuclear Research University ``MEPHI'', 115409 Moscow, Russia}

\maketitle

\begin{abstract}{{\bf graphene; pairing; massless fermions; superfluidity}}
The superconducting pairing of electrons in doped graphene due to in-plane and out-of-plane phonons is considered. It is shown that the structure of
the order parameter in the valley space substantially affects conditions of the pairing. Electron-hole pairing in graphene bilayer in the strong
coupling regime is also considered. Taking into account retardation of the screened Coulomb pairing potential shows a significant competition between
the electron-hole direct attraction and their repulsion due to virtual plasmons and single-particle excitations.
\end{abstract}

\section{Introduction}
Electrons in graphene, a two-dimensional form of carbon, can be described by a two-dimensional Dirac-type equation for massless particles near the
Fermi level (see Castro Neto \ea (2009) and references therein). Thereby, graphene offers a unique possibility to study effectively ultrarelativistic
charged particles in condensed matter phenomena (Katsnelson \ea 2006, Katsnelson \& Novoselov 2007) and particularly in collective phenomena (Lozovik
\ea 2008, Berman \ea 2008\textit{a,b}). Absence of mass for electrons make it possible to achieve new regimes of quantum many-particle systems
behavior in graphene. Therefore it is interesting to search for various superconducting and superfluid phases in graphene and graphene-based
structures, with their applications for dissipationless information transfer in nanoscale devices. In the present paper, we consider the
Bardeen-Cooper-Schrieffer-like (BCS-like) (Bardeen \ea 1957) phonon-mediated pairing of electrons in graphene, and the Coulomb pairing of electrons
and holes in graphene bilayer, taking into account the unusual electron dynamics.

Several possibilities for electron pairing phenomena in graphene were proposed. One possibility is the pairing of spontaneously created electrons in
the conduction band and holes in the valence band, leading to the excitonic insulator state (Khveshchenko 2001). The results of numerical simulations
in the recent papers (see, e.g., Drut \& L\"{a}hde (2008), Armour \ea (2009)) show that graphene can turn into excitonic insulator state while being
suspended in vacuum. Another possibility is electron-electron pairing in graphene layer, mediated either by phonons, plasmons (Uchoa \& Castro Neto
2007), or by Coulomb interaction, acting as attractive in certain channels (i.e. resonating valence bond mechanisms, proposed by Black-Schaffer \&
Doniach (2007), Honerkamp (2008), or anisotropic electron scattering near Van Hove singularity by Gonz\'{a}lez (2008)). The other two possible
mechanisms of establishing a coherent state in graphene are the proximity-induced superconductivity (Heersche \ea 2007, Beenakker 2008), and the
pairing of spatially separated electrons and holes in graphene bilayer (Lozovik \& Sokolik 2008\textit{a}, Min \ea 2008, Zhang \& Joglekar 2008),
analogous to electron-hole pairing in coupled quantum wells (Lozovik \& Yudson 1975,1976, Lozovik \& Poushnov 1997, Lozovik \& Berman 1997).

We consider electron-electron pairing due to optical in-plane phonons, represented by two pairs of doubly-degenerate modes (Piscanec \ea 2004, Basko
\& Aleiner 2008), and due to out-of-plane (flexural) acoustical and optical modes; the out-of-plane modes interact with electrons quadratically
(Mariani \& von Oppen 2008, Khveshchenko 2009). We demonstrate that each phonon mode in graphene provides a contribution to effective
electron-electron interaction, dependent on both its symmetry and the structure of the order parameter with respect to electron valleys. Estimates of
the coupling constants show that out-of-plane phonons do not cause a pairing with any observable critical temperatures, however the in-plane optical
phonons can lead to the pairing in heavily doped graphene.

Electron-hole pairing in graphene bilayer in the weak coupling regime is of BCS type, and affects only the conduction band of the electron-doped
graphene layer and the valence band of the hole-doped layer (Lozovik \& Sokolik 2008\textit{a}). On increase of the coupling strength, the pairing
become multi-band, involving also the valence band of the electron-doped layer and the conduction band of the hole-doped layer (Lozovik \& Sokolik
2009, Lozovik\& Sokolik 2010). Such ``ultrarelativistic'' regime of pairing occurs due to absence of localized pairs in graphene (Lozovik \& Sokolik
2008\textit{b}, Sabio \ea 2009) --- in contrast to usual systems of attracting fermions, where crossover to a gas of local pairs at strong coupling
occurs (Nozi\`{e}res \& Schmitt-Rink 1985).

The estimates of a critical temperature in graphene bilayer within the framework of one-band BCS model with taking into account static screening of
electron-hole interaction give unobservably small values (Kharitonov \& Efetov 2008\textit{a,b}). However the estimates using unscreened interaction,
or with the statically screened interaction, but within the multi-band model, provide much larger values of the critical temperature (Min \ea 2008,
Zhang \& Joglekar 2008, Lozovik \& Sokolik 2009). In this paper, we consider the strong coupling regime within the framework of multi-band model,
taking into account a dynamical screening of electron-hole interaction. The dynamical effects manifest themselves as virtual plasmons and
single-particle excitations, which contribution to the interaction is repulsive and thus competes with the ``direct'' Coulomb attraction.

The article is organized as follows. In Sec.~2 we derive and solve the gap equations for the phonon-mediated electron-electron pairing in graphene.
In Sec.~3 we study electron-hole pairing in graphene bilayer using Eliashberg-type equations. Sec.~3 is devoted to conclusions.

\section{Phonon-mediated electron-electron pairing in graphene}
Electrons in graphene populate two interpenetrating triangular lattices $A$ and $B$, composing the bipartite graphene lattice, and two ``valleys''
$\vec{K}$ and $\vec{K}'=-\vec{K}$ in momentum space, therefore it is convenient to describe electrons by the effective four-component wave function
(Castro Neto \ea 2009). Analogously to Gusynin \ea (2007), we introduce the four-component electron destruction operator
$\Psi_{\vec{p}}=(a_{\vec{K}+\vec{p}},b_{\vec{K}+\vec{p}},b_{\vec{K}'+\vec{p}},a_{\vec{K}'+\vec{p}})^T$, where the operators $a_{\vec{p}}$ and
$b_{\vec{p}}$ correspond to sublattices $A$ and $B$ respectively. In the Heisenberg representation, $\Psi_{\vec{p}}$ evolves according to the
Dirac-type equation:
\begin{eqnarray}
p_\mu\gamma^\mu\Psi_{\vec{p}}=0,\qquad\mu=0,1,2.\label{Dirac}
\end{eqnarray}
The ``covariant'' coordinates $p^0=(i/v_\mathrm{F})(\partial/\partial t)$, $p^{1,2}=p_{x,y}$, $p_\mu=\{p^0,-p^1,-p^2\}$ are used, where
$v_\mathrm{F}\approx10^6\,\mbox{m/s}$ is the Fermi velocity. The gamma matrices are in the Weyl representation:
\begin{eqnarray}
\gamma^0=\left(\begin{array}{cc}0&I\\I&0\end{array}\right),\quad\vec{\gamma}=\left(\begin{array}{cc}0&-\vec{\sigma}\\
\vec{\sigma}&0\end{array}\right),\quad\gamma^5=i\gamma^0\gamma^1\gamma^2\gamma^3=\left(\begin{array}{cc}I&0\\0&-I\end{array}\right).\nonumber
\end{eqnarray}

\begin{figure}[t]
\begin{center}
\resizebox{0.9\textwidth}{!}{\includegraphics{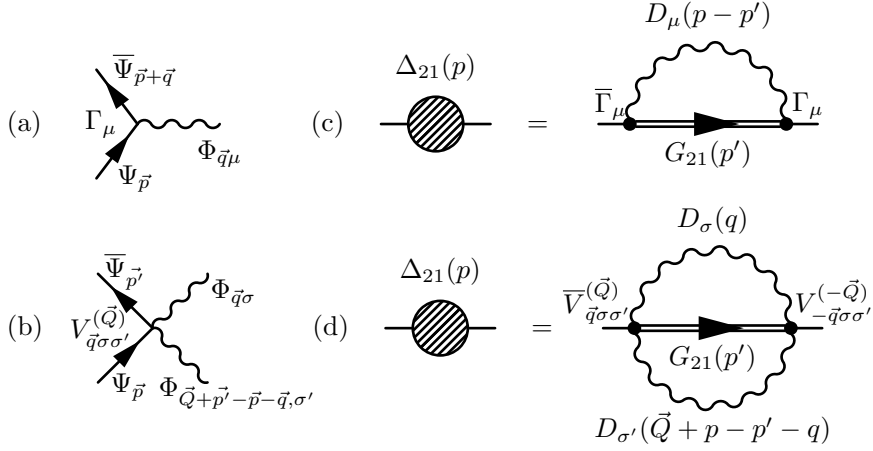}}
\end{center}
\caption{\label{Fig1} Diagrammatic representations of electron-phonon coupling Hamiltonians (\ref{H_int_1}) and (\ref{H_int_3}) for linear (a) and
quadratic (b) couplings respectively. The corresponding self-consistent gap equations (\ref{Sc1}) and (\ref{Sc2}) are shown in (c) and (d)
respectively.}
\end{figure}

The Hamiltonian of linear electron-phonon coupling can be written in the general form (see Fig.~\ref{Fig1}(a)):
\begin{eqnarray}
H^\mathrm{(lin)}_\mathrm{el-ph}=\frac1{\sqrt{S}}\sum_{\vec{p}\vec{q}\mu}
g^{(\mu)}_{\vec{p}\vec{q}}\overline\Psi_{\vec{p}+\vec{q}}\Gamma_\mu\Psi_{\vec{p}}\Phi_{\vec{q}\mu}.\label{H_int_1}
\end{eqnarray}
Here $g^{(\mu)}_{\vec{p}\vec{q}}$ and $\Gamma_\mu$ are coupling amplitude and interaction vertex for the $\mu$-th phonon mode,
$\Phi_{\vec{q}\mu}=c_{\vec{q}\mu}+c^+_{-\vec{q}\mu}$, where $c_{\vec{q}\mu}$ is the phonon destruction operator;
$\overline\Psi_{\vec{p}}=\Psi_{\vec{p}}^+\gamma^0$ is the Dirac-conjugated spinor and $S$ is the system area.

We take into account two pairs of degenerate in-plane optical phonon modes, most strongly coupled to graphene electrons (Piscanec \ea 2004, Basko \&
Aleiner 2008): $A_1$ and $B_1$ modes (we denote them by $\mu=1,2$) with the momentum $\vec{q}=\pm\vec{K}$ and energy
$\omega_{\vec{K}}\approx0.170\,\mbox{eV}$, and $E_{2x}$ and $E_{2y}$ modes ($\mu=3,4$) with $\vec{q}=\Gamma$, $\omega_\Gamma\approx0.196\,\mbox{eV}$.
The coupling constants (Piscanec \ea 2004) and interaction vertices (Basko \& Aleiner 2008) for these modes are:
$g^{(1,2)}_{\vec{p}\vec{q}}\approx1.34\,\mbox{eV}\cdot\mbox{\AA}$, $g^{(3,4)}_{\vec{p}\vec{q}}\approx0.86\,\mbox{eV}\cdot\mbox{\AA}$; $\Gamma_1=I$,
$\Gamma_2=i\gamma^5$, $\Gamma_3=-\gamma^5\gamma^2$, $\Gamma_4=-\gamma^5\gamma^1$.

The Hamiltonian of quadratic interaction of graphene electrons with out-of-plane phonons is more complicated. The major part of the interaction,
resulting from the deformation potential (Suzuura \& Ando 2002, Mariani \& von Oppen 2008), can be written in the form:
\begin{eqnarray}
H^\mathrm{(quadr)}_\mathrm{el-ph}=\frac1S\sum_{\vec{p}\vec{p'}\vec{q}}\sum_{\sigma\sigma'}\sum_{j=1}^3
\left[a^+_{\vec{p'}}a_{\vec{p}}+b^+_{\vec{p'}}b_{\vec{p}}\,e^{i(\vec{p}-\vec{p'})\vec{d}_j}\right]\Phi_{\vec{q}\sigma}\Phi_{\vec{q'}\sigma'}\nonumber\\
\times\frac{\sqrt3g_1\left[e^{i\vec{q}\vec{d}_j}\varepsilon^B_{\vec{q}\sigma}-\varepsilon^A_{\vec{q}\sigma}\right]
\left[e^{i\vec{q'}\vec{d}_j}\varepsilon^B_{\vec{q'}\sigma'}-\varepsilon^A_{\vec{q'}\sigma'}\right]}
{4M\sqrt{\omega_{\vec{q}\sigma}\omega_{\vec{q'}\sigma'}}},\label{H_int_2}
\end{eqnarray}
where $\vec{q'}=\vec{p'}-\vec{p}-\vec{q}$, $M$ is the carbon atom mass, $g_1\approx20-30\,\mbox{eV}$ and $\vec{d}_j$ ($j=1,2,3$) are the vectors
connecting an atom from the $A$ sublattice with its nearest neighbors. The summation over $\vec{p}$, $\vec{p'}$ and $\vec{q}$ in (\ref{H_int_2}) is
performed over the first Brillouin zone of graphene; $\varepsilon^{A,B}_{\vec{q}\sigma}$ and $\omega_{\vec{q}\sigma}$ are polarizations and
frequencies of acoustical ($\sigma=1$) and optical ($\sigma=2$) out-of-plane phonons branches.

Rewriting (\ref{H_int_2}) in the four-component spinor notation requires splitting of the summation over electron momentums $\vec{p}$, $\vec{p'}$
among the valleys $\pm\vec{K}$. The result is (see Fig.~\ref{Fig1}(b); see also Lozovik \& Sokolik (in press) for details):
\begin{eqnarray}
H^\mathrm{(quadr)}_\mathrm{el-ph}=\frac1S\sum_{\vec{p}\vec{p'}\vec{q}}\sum_{\sigma\sigma'}\sum_{\vec{Q}=\Gamma,\pm\vec{K}}
\overline\Psi_{\vec{p'}}V^{(\vec{Q})}_{\vec{q}\sigma\sigma'}\Psi_{\vec{p}}
\Phi_{\vec{q}\sigma}\Phi_{\vec{Q}+\vec{p'}-\vec{p}-\vec{q},\sigma'},\label{H_int_3}
\end{eqnarray}
\begin{eqnarray}
V^{(\vec{Q})}_{\vec{q}\sigma\sigma'}=\frac{\sqrt3g_1}{8M\sqrt{\omega_{\vec{q}\sigma}\omega_{\vec{q'}\sigma'}}}
\left\{\delta_{\vec{Q}\Gamma}\gamma^1\gamma^5+\delta_{\vec{Q}\vec{K}}(1+\gamma^5)+\delta_{\vec{Q},-\vec{K}}(1-\gamma^5)\right\}
i\gamma^2\nonumber\\
\times\sum_{j=1}^3\left[\gamma^3+\gamma^0+(\gamma^3-\gamma^0)e^{-i\vec{Q}\vec{d}_j}\right]
\left[e^{i\vec{q}\vec{d}_j}\varepsilon^B_{\vec{q}\sigma}-\varepsilon^A_{\vec{q}\sigma}\right]
\left[e^{i\vec{q'}\vec{d}_j}\varepsilon^B_{\vec{q'}\sigma'}-\varepsilon^A_{\vec{q'}\sigma'}\right],\nonumber
\end{eqnarray}
where $\vec{q'}=\vec{Q}-\vec{q}$.

Similarly to Pisarski \& Rischke (1999), we describe the pairing by the set of matrix Green functions in Matsubara representation
$G_{ij}(\vec{p},\tau)=-\langle T\Psi^{(i)}_{\vec{p}}(\tau)\overline\Psi^{(j)}_{\vec{p}}(0)\rangle$, where $\Psi^{(1)}_{\vec{p}}=\Psi_{\vec{p}}$,
$\Psi^{(2)}_{\vec{p}}=\Psi_{\mathrm{C}\vec{p}}\equiv C\overline\Psi^T_{-\vec{p}}$ is the charge-conjugated spinor and $C=i\gamma^2\gamma^0$ is the
charge-conjugation matrix. The anomalous Green functions $G_{12}$ and $G_{21}$ are responsible for a Cooper pair condensate. The Gor'kov equations,
describing the pairing in the mean-field approximation, are:
\begin{eqnarray}
G_{ij}(p)=\delta_{ij}G^{(0)}_i(p)+G^{(0)}_i(p)\Delta_{i,3-i}(p)G_{3-i,j}(p),\label{gor}
\end{eqnarray}
where $p=\{p_0=i\pi T(2k+1),\vec{p}\}$, $G^{(0)}_{1,2}(p)=[\gamma^0(p_0\pm\mu)-v_\mathrm{F}\vec{\gamma}\vec{p}]^{-1}$ are the free particle Green
functions, following from (\ref{Dirac}) and $\mu$ is the chemical potential in graphene.

The expressions for the anomalous self-energy $\Delta_{21}$ (the other component $\Delta_{12}=\gamma^0\Delta_{21}^+\gamma^0$) in (\ref{gor}) for the
cases of linear (\ref{H_int_1}) and quadratic (\ref{H_int_3}) electron-phonon interaction Hamiltonians are, respectively (see Fig.~\ref{Fig1}(c,d)),
\begin{eqnarray}
\Delta_{21}(p)=-\frac{T}S\sum_{p'\mu}g_\mu^2D_\mu(p-p')\overline\Gamma_\mu G_{21}(p')\Gamma_\mu,\label{Sc1}
\end{eqnarray}
\begin{eqnarray}
\Delta_{21}(p)=\frac{2T^2}{S^2}\sum_{p'q}\sum_{\vec{Q}\sigma\sigma'}D_\sigma(q)D_{\sigma'}(\vec{Q}+p-p'-q)
\overline{V}^{(\vec{Q})}_{\vec{q}\sigma\sigma'}G_{21}(p')V^{(-\vec{Q})}_{-\vec{q}\sigma\sigma'}.\label{Sc2}
\end{eqnarray}
Here the charge-conjugated vertices $\overline\Gamma_\mu=C^{-1}\Gamma_\mu^TC$ and
$\overline{V}^{(\vec{Q})}_{\vec{q}\sigma\sigma'}=C^{-1}V^{(\vec{Q})T}_{\vec{q}\sigma\sigma'}C$ are introduced, and
$D_\mu(q)=2\omega_{\vec{q}\mu}/(q_0^2-\omega_{\vec{q}\mu}^2)$ is the phonon Green function.

To solve the Gor'kov equations (\ref{gor}), we assume that the pairing is diagonal with respect to the conduction and valence bands, but suppose,
that the structure of the order parameter with respect to electron valleys is parameterized by arbitrary $SU(2)$ matrix. The band-diagonal order
parameter can be represented as a decomposition over projection operators $\mathcal{P}_\pm(\hat{p})=(1\pm\gamma^0\vec{\gamma}\hat{p})/2$ on
conduction and valence bands, where $\hat{p}=\vec{p}/|\vec{p}|$ (Pisarski \& Rischke 1999). Further, rotation of the order parameter in the valley
space can be performed by means of three generators $T_1=\gamma^5$, $T_2=\gamma^3\gamma^5$ and $T_3=i\gamma^3$, obeying the algebra of the Pauli
matrices (Gusynin \ea 2007). Thus, the explicit form of $\Delta_{21}$ is:
\begin{eqnarray}
\Delta_{21}(p)=e^{i\vec{v}\vec{T}}\left[\Delta_+(p)\mathcal{P}_+(\hat{p})+\Delta_-(p)\mathcal{P}_-(\hat{p})\right],\label{Delta_21}
\end{eqnarray}
where $\Delta_\pm(p)$ are the gaps in conduction and valence bands, and the three-dimensi\-onal vector $\vec{v}$ defines the valley structure of the
order parameter.

Substituting (\ref{Delta_21}) in (\ref{Sc1}) and (\ref{Sc2}), we can derive the system of two coupled gap equations for $\Delta_\pm(p)$, having the
form:
\begin{eqnarray}
\Delta_\alpha(p)=-\frac{T}S\sum_{p'\beta}\frac{\Delta_\beta(p')}{p_0'^2-E^2_\beta(p')}\,\Lambda^\mu_{\alpha\beta}(p,p';\vec{v}),\label{Sc3}
\end{eqnarray}
where $E_\pm(p)=\sqrt{(v_\mathrm{F}|\vec{p}|\mp\mu)^2+\Delta^2_\pm(p)}$ are the energies of Bogolyubov excitations in conduction and valence bands.
The effective interband ($\alpha=\beta$) and intraband ($\alpha=-\beta$) interactions in the cases of linear and quadratic electron-phonon couplings
are, respectively,
\begin{eqnarray}
\Lambda^\mu_{\alpha\beta}(p,p';\vec{v})=\frac12\sum_\mu g_\mu^2D_\mu(p-p')\,\mathrm{Sp}\left[\mathcal{P}_\alpha(\hat{p})e^{-i\vec{v}\vec{T}}
\overline{\Gamma}_\mu\gamma^0e^{i\vec{v}\vec{T}}\mathcal{P}_\beta(\hat{p}')\gamma^0\Gamma_\mu\right],\label{V1}
\end{eqnarray}
\begin{eqnarray}
\Lambda^\mu_{\alpha\beta}(p,p';\vec{v})=-\frac{T}S\sum_{\vec{q}\vec{Q}}\sum_{\sigma\sigma'}D_\sigma(q)D_{\sigma'}(\vec{Q}+p-p'-q)\nonumber\\
\times\mathrm{Sp}\left[\mathcal{P}_\alpha(\hat{p})e^{-i\vec{v}\vec{T}}\overline{V}^{(\vec{Q})}_{\vec{q}\sigma\sigma'}
\gamma^0e^{i\vec{v}\vec{T}}\mathcal{P}_\beta(\hat{p}')\gamma^0V^{(-\vec{Q})}_{-\vec{q}\sigma\sigma'}\right].\label{V2}
\end{eqnarray}

\begin{figure}[t]
\begin{center}
\resizebox{0.6\textwidth}{!}{\includegraphics{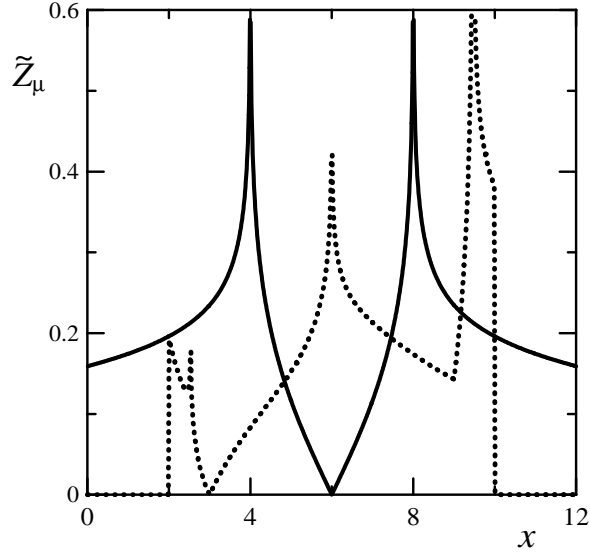}}
\end{center}
\caption{\label{Fig2} Partial dimensionless Eliashberg functions for out-of-plane phonons. Contributions of the two-phonon processes, leaving the
electron in its initial valley ($\tilde{Z}_\Gamma$, solid line), and flipping it into the opposite valley ($\tilde{Z}_{\vec{K}}$, dotted line) are
shown.}
\end{figure}

We can find analytical solutions of (\ref{Sc3})--(\ref{V2}) in the regime of high doping of graphene, when $\mu$ is greater than the characteristic
phonon frequencies. Moreover, high doping facilitates the pairing due to larger density of states at the Fermi level $\mathcal{N}=\mu/2\pi
v_\mathrm{F}^2$. In this case the pairing is effectively one-band, and the Eliashberg equations at $T=0$, following from (\ref{Sc3}), can be derived
by setting $\Delta_+(p)=\Delta\Theta(\omega_0-p_0)$, $\Delta_-=0$ (similarly to Lozovik \ea 2010):
\begin{eqnarray}
1=2\int\limits_0^{\omega_0}\frac{d\omega}{\sqrt{\omega^2-\Delta^2}}\int\limits_0^\infty
d\nu\frac{\alpha^2_{\vec{v}}(\nu)F(\nu)}{\omega+\nu}.\label{Sc4}
\end{eqnarray}
Here $\omega_0$ is a cutoff frequency of the order of phonon frequencies, and the Eliashberg functions $\alpha^2_{\vec{v}}(\nu)F(\nu)$ for in-plane
and out-of-plane phonons respectively can be represented as:
\begin{eqnarray}
\alpha^2_{\vec{v}}(\nu)F(\nu)=\mathcal{N}g_\Gamma^2\delta(\nu-\omega_\Gamma)R^{(1)}_\Gamma(\vec{v})+
\mathcal{N}g_{\vec{K}}^2\delta(\nu-\omega_{\vec{K}})R^{(1)}_{\vec{K}}(\vec{v}),\label{el1}
\end{eqnarray}
\begin{eqnarray}
\alpha^2_{\vec{v}}(\nu)F(\nu)=Z_\Gamma(\nu)R^{(2)}_\Gamma(\vec{v})+Z_{\vec{K}}(\nu)R^{(2)}_{\vec{K}}(\vec{v}).\label{el2}
\end{eqnarray}
The partial Eliashberg functions of out-of-plane phonons can be calculated within the simple phonon model (see Lozovik \& Sokolik (in press)) and
reduced to dimensionless functions: $Z_\mu(\nu)=(81\mathcal{N}g_1^2/2a^2M^2\omega_{\Gamma2}^3)\tilde{Z}_\mu(x)$, where $x=6\nu/\omega_{\Gamma2}$,
$\omega_{\Gamma2}\approx0.11\,\mbox{eV}$. The functions $\tilde{Z}_\mu(x)$ are shown in Fig.~\ref{Fig2}.

Eqs. (\ref{el1})--(\ref{el2}) show that the Eliashberg functions consist of two parts, corresponding to phonon-mediated electron-electron interaction
processes, leaving both electrons in their valleys ($\Gamma$-terms), and flipping both electrons into opposite valleys ($\vec{K}$-terms). The factors
$R^{(1,2)}_{\Gamma,\vec{K}}(\vec{v})$, dependent on the valley structure of the order parameter, determine the signs and amplitudes of these
contributions:
\begin{eqnarray}
R^{(1)}_\Gamma(\vec{v})=-\cos^2v+(1-2\hat{v}_1^2)\sin^2v,\quad
R^{(1)}_{\vec{K}}(\vec{v})=-\cos^2v+\hat{v}_1^2\sin^2v,\nonumber\\
R^{(2)}_\Gamma(\vec{v})=1,\quad R^{(2)}_{\vec{K}}(\vec{v})=(-\hat{v}_2^2+\hat{v}_3^2)\sin^2v.\label{fact}
\end{eqnarray}
All the factors (\ref{fact}) except $R_\Gamma^{(2)}$ can vary in the range from $-1$ (the effective phonon-mediated repulsion) to $+1$ (the effective
attraction).

The system under pairing conditions should prefer the valley structure $\vec{v}$ of the order parameter, which provides the maximal gap and thus the
maximal $\alpha^2_{\vec{v}}(\nu)F(\nu)$. For in-plane phonons, the gap, found from (\ref{Sc4})--(\ref{el1}), is
$\Delta\sim\omega_{\Gamma,\vec{K}}\exp\{-1/\lambda\}$, where the effective coupling constant
$\lambda=2\int_0^\infty(d\nu/\nu)\alpha^2_{\vec{v}}(\nu)F(\nu)$ consists of the partial coupling constants: $\lambda=\lambda_\Gamma
R_\Gamma(\vec{v})+\lambda_{\vec{K}}R_{\vec{K}}(\vec{v})$, $\lambda_\mu=2\mathcal{N}g_\mu^2/\omega_\mu$. When $\lambda_{\vec{K}}>2\lambda_\Gamma$, the
preferable pairing structure is $\vec{v}=\{\pi/2,0,0\}$: in this case $R_{\vec{K}}=1$ (scalar and pseudoscalar phonons give rise to effective
electron-electron attraction), $R_\Gamma=-1$ (pseudovector phonons cause repulsion). At $\lambda_{\vec{K}}<2\lambda_\Gamma$, we have
$\vec{v}=\{0,(\pi/2)\cos\varphi,(\pi/2)\sin\varphi\}$, and $R_\Gamma=1$ (pseudovector phonons cause attraction), $R_{\vec{K}}=0$ (contributions from
scalar and pseudoscalar modes cancel each other). Actually, at high values of dielectric permittivity of surrounding medium,
$\lambda_{\vec{K}}<2\lambda_\Gamma$; at lower permittivity, the Coulomb interaction renormalizes $\lambda_{\vec{K}}$ towards higher values, so the
relation $\lambda_{\vec{K}}>2\lambda_\Gamma$ can be satisfied (Basko \& Aleiner 2008).

For out-of-plane phonons, the preferable valley structure of the order parameter is $\vec{v}=\{0,0,\pi/2\}$, when $R_\Gamma=R_{\vec{K}}=1$. Numerical
estimates of the coupling constant for out-of-plane phonons show very small values ($10^{-3}$ by the order of magnitude), thus out-of-plane phonons
cannot provide any observable electron pairing in graphene. However, in-plane optical phonons can provide observable pairing at heavy doping of
graphene.

\section{Electron-hole pairing in graphene bilayer}
The pairing of spatially separated electrons and holes in graphene bilayer occurs due to Coulomb attraction. This longitudinally-vectorial
interaction is described by the vertex $\Gamma=\gamma^0$ within the framework of matrix diagrammatic technique, employed in the previous section. Any
interaction with such vertex lead to effective electron-electron interaction, independent on the valley structure of the order parameter. Under
band-diagonal pairing, the system of self-consistent equations for the conduction- and valence-band gap functions $\Delta_\pm(p)$ is similar to
(\ref{Sc3}):
\begin{eqnarray}
\Delta_\alpha(p)=-\frac{T}S\sum_{p'\beta}\frac{1+\alpha\beta\hat{p}\hat{p}'}2V(|\vec{p}-\vec{p}'|,p_0-p_0')\,F_\beta(p'),\label{BL_sc1}
\end{eqnarray}
where $F_\beta(p')=\Delta_\beta(p')/[p_0'^2-E_\beta^2(p')]$ is the anomalous Green function and $V(q,\omega)$ is the dynamically screened
electron-electron interaction.

The system (\ref{BL_sc1}) can be solved in the spirit of BCS theory (Bardeen \ea 1957), i.e. in the static approximation, when one puts $\omega=0$ in
$V(q,\omega)$ and assume, that $\Delta_\pm(\vec{p},\omega)$ do not depend on $\omega$ and are non-zero in some range of $p$, corresponding to
neighborhood of the Fermi surface. Such calculations (Lozovik \& Sokolik 2009) showed that in multi-band pairing regime the gap depends exponentially
on energy width of the pairing region and thus can be very large. Here we go beyond the static approximation and take into account the frequency
dependence of the screened interaction.

The pairing interaction $V(q,\omega)$ can be calculated in the random phase approximation, well-justified in graphene bilayer due to large number of
fermionic flavors, equal to 8 (Kharitonov \& Efetov 2008\textit{a,b}; see also Apenko \ea 1982):
\begin{eqnarray}
V(q,\omega)=\frac{v_qe^{-qD}}{1-2v_q\Pi(q,\omega)+v_q^2\Pi^2(q,\omega)(1-e^{-2qD})},\label{int1}
\end{eqnarray}
where $v_q=2\pi e^2/\varepsilon q$ is the bare Coulomb interaction, $\varepsilon$ is a dielectric permittivity of surrounding medium, $\Pi(q,\omega)$
is a polarization operator of each graphene layer. Hereafter we consider the case of small interlayer distance $D$, when $p_\mathrm{F}D\ll1$,
$p_\mathrm{F}=\mu/v_\mathrm{F}$ is the Fermi momentum. At $\omega=0$, (\ref{int1}) reduces to the statically screened interaction, equal in
dimensionless form to $\mathcal{N}V(q,0)=r_\mathrm{s}/(q/p_\mathrm{F}+8r_\mathrm{s})$, where $r_\mathrm{s}=e^2/\varepsilon
v_\mathrm{F}\approx2.19/\varepsilon$ determines the coupling strength. There exist two plasmon branches in the system, corresponding to zeros of
denominator of (\ref{int1}): the lower branch with the dispersion $\omega_-(q)\approx v_\mathrm{F}q$, and the upper branch with the square-root
dispersion $\omega_+(q)\approx2\mu\sqrt{r_\mathrm{s}(q/p_\mathrm{F})}$ at small $q$ and almost linear dispersion at large $q$. When
$\omega_\pm(q)+v_\mathrm{F}q>2\mu$, the plasmons in graphene acquire a finite lifetime due to interband transitions (Wunsch \ea 2006, Hwang \& Das
Sarma 2007). At $\omega\rightarrow\infty$, the potential $V(q,\omega)$ becomes unscreened: $\mathcal{N}V(q,\infty)=r_\mathrm{s}/(q/p_\mathrm{F})$.

Using the spectral representations of $F_\beta(p')$ and $V(q)$ in (\ref{BL_sc1}) and summing over $p_0'$ at $T=0$, we get the Eliashberg-type
equations:
\begin{eqnarray}
\Delta_\alpha(\vec{p},\omega)=-\sum_\beta\int\frac{d\vec{p'}}{(2\pi)^2}\frac{1+\alpha\beta\hat{p}\hat{p}'}2
\int\limits_0^\infty\frac{d\omega'}\pi\,\mathrm{Im}F_\beta(\vec{p'},\omega')
\left\{v_{\vec{p}-\vec{p'}}+\int\limits_0^\infty\frac{d\nu}\pi\right.\nonumber\\
\left.\vphantom{\int\limits_0^\infty}\times\mathrm{Im}V(\vec{p}-\vec{p'},\omega-\omega'+i\delta)
\left(\frac1{\omega'+\nu+\omega+i\delta}-\frac1{\omega'+\nu-\omega-i\delta}\right)\right\}.\label{BL_sc2}
\end{eqnarray}
We assume that $\Delta(\vec{p},\omega)$ is real (it is well-justified near $\omega=0$), therefore
$\mathrm{Im}F_\beta(p')=-\pi\delta(\omega'-E_\beta(p'))\Delta_\beta(p')/2E_\beta(p')$. Then, we assume that the argument of the $\delta$-function in
this expression vanishes at some unambiguous $\omega'=\tilde\omega(\vec{p'})$. This allows us to handle only ``on-shell'' gap functions and
Bogolyubov energies: $\Delta_\alpha(|\vec{p}|)\equiv\Delta_\alpha(\vec{p},\tilde\omega(\vec{p}))$, $E_\alpha(|\vec{p}|)\equiv
E_\alpha(\vec{p},\tilde\omega(\vec{p}))$. Rewriting (\ref{BL_sc2}) in terms of the on-shell quantities and assuming $\alpha=+1$, $p=p_\mathrm{F}$ in
its left-hand side, we get:
\begin{eqnarray}
\Delta_+(p_\mathrm{F})=\sum_\beta\int\frac{d\vec{p'}}{(2\pi)^2}\frac{1+\beta\hat{p}\hat{p}'}2\frac{\Delta_\beta(p')}{2E_\beta(p')}\nonumber\\
\times\left\{v_{\vec{p}-\vec{p'}}+\frac2\pi\int\limits_0^\infty\mathrm{Im}V(\vec{p}-\vec{p'},\nu+i\delta)\frac{d\nu}{E_\beta(p')+\nu}\right\}.
\label{BL_sc3}
\end{eqnarray}
Here we also neglected $\omega=\Delta_+(p_\mathrm{F})$ in the second term in the braces.

To demonstrate the influence of dynamical screening of $V(q,\omega)$ on the gap value, we will simplify the equation (\ref{BL_sc3}) further. Firstly,
we suppose that the both on-shell gap functions $\Delta_\beta(p)$ are equal to each other; this seems plausible at large $r_\mathrm{s}$, as discussed
by Lozovik \& Sokolik 2010, and allows us to neglect the angular factor $\hat{p}\hat{p}'$ in (\ref{BL_sc3}). Secondly, we assume
$\Delta_\pm(p)=\Delta f(p)$, where $\Delta=\Delta_+(p_\mathrm{F})$ is the gap at the Fermi surface and $f(p)$ is some trial function, equal to 1 at
$p=p_\mathrm{F}$ and having the asymptotics $f(p)\propto1/p$ at $p\rightarrow\infty$, caused by the leading contribution of the unscreened Coulomb
interaction in (\ref{BL_sc2}). For computational purposes, it is also convenient to single out the contribution of undamped plasmons of the higher
branch as: $\mathrm{Im}V(q,\nu+i\delta)=\mathrm{Im}V(q,\nu)-\pi\delta[\nu-\omega_+(q)]A(q)$. Here the spectral weight of the higher branch plasmons
is $A(q)=-v_q/2[\partial\Pi(q,\omega)/\partial\omega]|_{\omega=\omega_+(q)}$ when $q+\omega_+(q)<2\mu$ and zero otherwise. Thus, the equation
(\ref{BL_sc3}) for the gap reduces to:
\begin{eqnarray}
1=\frac12\sum_\beta\int\frac{d\vec{p}}{(2\pi)^2}\frac{\Delta_\beta(p)}{2E_\beta(p)}\left\{v_{\vec{p}-\vec{p}_\mathrm{F}}
-\frac{2A(\vec{p}-\vec{p}_\mathrm{F})}{E_\beta(p)+\omega_+(\vec{p}-\vec{p}_\mathrm{F})}+\vphantom{\int\limits_0^\infty}\right.\nonumber\\
\left.+\frac2\pi\int\limits_0^\infty\mathrm{Im}V(\vec{p}-\vec{p}_\mathrm{F},\nu)\frac{d\nu}{E_\beta(p)+\nu}\right\}.\label{BL_sc4}
\end{eqnarray}

\begin{figure}[t]
\begin{center}
\resizebox{0.6\textwidth}{!}{\includegraphics{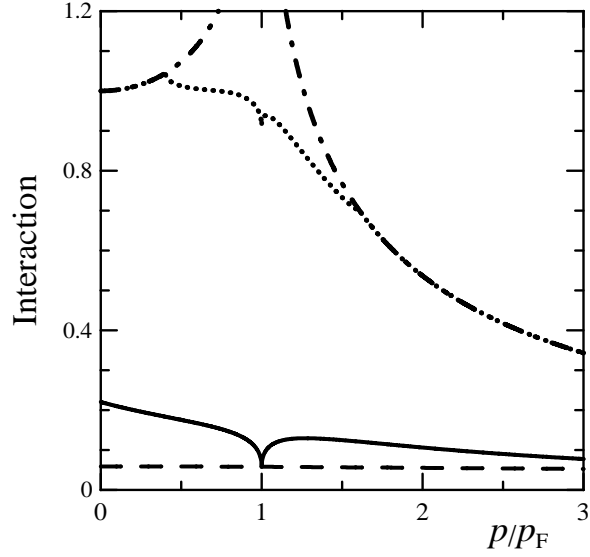}}
\end{center}
\caption{\label{Fig3} On-shell interaction, represented by the expression in the braces of (\ref{BL_sc4}), in the intraband channel ($\beta=+1$) at
$r_\mathrm{s}=2$. Solid line: full on-shell interaction, dash-dotted line: unscreened Coulomb interaction only, dotted line: unscreened Coulomb
interaction with the contribution of undamped plasmons, dashed line: statically screened interaction.}
\end{figure}

The expression in braces in (\ref{BL_sc4}) is the effective on-shell interaction, which incorporates the effects of dynamical screening of the
pairing interaction. It differs from the statically screened interaction $V(q,\omega)$, employed by Lozovik \& Sokolik 2009, Lozovik \& Sokolik 2010
and is naturally divided into three contributions (see Fig.~\ref{Fig3}): a) attractive unscreened Coulomb interaction, b) repulsive contribution due
to virtual undamped higher-branch plasmons, c) repulsive contribution due to virtual damped higher-branch plasmons and single-particle intra- and
interband excitation continuums at $\omega<v_\mathrm{F}q$ and $\omega+v_\mathrm{F}q>2\mu$ respectively (Wunsch \ea 2006, Hwang \& Das Sarma 2007).
However, strictly on the Fermi surface (at $\beta=+1$, $p=p_\mathrm{F}$) the effective on-shell interaction coincides with the statically-screened
one.

\begin{figure}[t]
\begin{center}
\resizebox{0.6\textwidth}{!}{\includegraphics{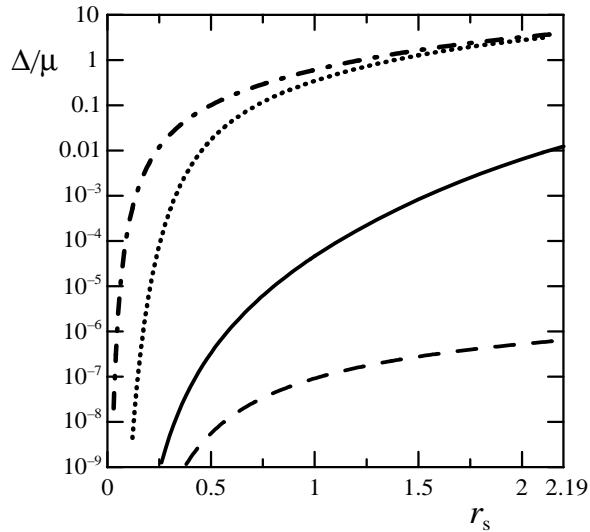}}
\end{center}
\caption{\label{Fig4} Values of the gap at the Fermi surface $\Delta$, normalized on the chemical potential $\mu$, as functions of $r_\mathrm{s}$,
calculated with taking into account various contributions to the effective on-shell interaction. Solid line: full on-shell interaction, dash-dotted
line: unscreened Coulomb interaction only, dotted line: unscreened Coulomb interaction with the contribution of undamped plasmons, dashed line:
statically screened interaction.}
\end{figure}

Assuming that the characteristic momentum, at which the on-shell gap $\Delta_\pm(p)$ decays, is of the order of the Fermi momentum and employing
$f(p)=p_\mathrm{F}/(|p-p_\mathrm{F}|+p_\mathrm{F})$ as a trial function, we can study the influence of dynamical screening of the pairing interaction
on the gap value $\Delta$. Fig.~\ref{Fig4} shows the results of the numerical solution of (\ref{BL_sc4}) with taking into account various
contributions to the effective on-shell interaction. Taking into account only the unscreened Coulomb attraction results in huge gap values, reported
previously by Min \ea (2008), Zhang \& Joglekar (2008). The addition of the undamped plasmons repulsive contribution eliminates the logarithmic
singularity of the effective interaction on the Fermi momentum (Fig.~\ref{Fig3}), but do not change the gap values essentially at large
$r_\mathrm{s}$. Finally, the contribution of the damped plasmons and and single-particle excitations lowers the gap down to the values of the order
of $0.01\mu$ at maximum (several Kelvins at $\mu\sim0.1\,\mbox{eV}$).

When the on-shell interaction is replaced by the statically screened one $V(q,0)$, the gap is unobservably small, if we use the trial function
$f(p)$, spread over the momentums of the order of $p_\mathrm{F}$; this is in argeement with the BCS-type estimates by Kharitonov \& Efetov
(2008\textit{a,b}). However, as shown by Lozovik \& Sokolik (2009), the pairing tend to occupy much larger region of the momentums of the order of
$8r_\mathrm{s}p_\mathrm{F}$ and results in large gap values, if we use the statically screened potential as the pairing potential. In our case, when
we take into account the dynamical effects and naturally assume that the on-shell gaps $\Delta_\pm(p)$ decay at $p\sim p_\mathrm{F}$, the gap turns
out to take the values, several orders of magnitude smaller than these with unscreened Coulomb interaction, but, at the same time, several orders of
magnitude larger than the BCS-type estimates.

\section{Conclusions}
We have considered electron-electron pairing in graphene, mediated by in-plane and out-of-plane phonons, and electron-hole pairing in graphene
bilayer, mediated by the screened Coulomb interaction. In both cases we consider the generally multi-band pairing with the $s$-wave order parameter,
diagonal with respect to valence and conduction bands of paired particles. Moreover, we take into account the frequency dependence of pairing
interaction in both cases, deriving and solving two-band Eliashberg-type equations.

The consideration of phonon-mediated pairing is performed by resolution of electron-phonon interaction with respect to sublattice and valley degrees
of freedom of electrons in graphene and taking into account a possibility of different structures of the order parameter in valley space. We
demonstrate that contributions of different phonon modes in graphene to effective electron-electron interaction, entering the Eliashberg-type
equations, depend both on symmetries of these modes and on the structure of the order parameter in valley space.

The coupling of graphene electrons with out-of-plane (flexural) phonon modes is quadratic and leads to unusual form of effective electron-electron
phonon-mediated interaction, which includes integration on frequency and integration on momentum over the whole Brillouin zone within the phonon
loop. The estimates of effective coupling constants show that the pairing due to in-plane phonon modes can occur at high doping of graphene, while
the pairing due to out-of-plane phonons does not occur at observable temperatures.

Eliashberg-type equations for electron-hole pairing in graphene bilayer, written in the present paper in terms of the ``on-shell'' gap functions,
allow to estimate the role of dynamical effects. The effective on-shell dynamically screened interaction, entering the gap equations, can be
represented as a sum of attractive unscreened Coulomb interaction, repulsive contribution due to undamped virtual plasmons and combined repulsive
contribution of the damped plasmons and of the continuum of single-particle excitations.

The unscreened Coulomb interaction on its own provides large values of the gap, which are only slightly reduced with taking into account the undamped
plasmons. Inclusion of the damped plasmons and single-particle excitations lower down the gap by several orders of magnitude. This result
demonstrates the significant competition between the bare Coulomb attraction and virtual excitations in the system, responsible for its dynamical
screening. However, the estimates of the gap, calculated with taking into account the full on-shell potential, are by several orders of magnitude
larger than the BCS-type estimates and can reach several Kelvins at strong coupling.

\end{document}